\documentstyle[12pt,aaspp4,tighten]{article}

\newcommand{\msun}{$M/M_{\odot}\,$}

\begin{document}

\title{METAL RICH RR LYRAE VARIABLES: I. THE EVOLUTIONARY SCENARIO}

\author{Giuseppe Bono}
\affil{Osservatorio Astronomico di Trieste, Via G.B. Tiepolo 11,
34131 Trieste, Italy; bono@oat.ts.astro.it}

\author{Filippina Caputo}
\affil{Osservatorio Astronomico di Capodimonte, Via Moiariello 16,
80131 Napoli, Italy; caputo@astrna.na.astro.it}

\author{Santi Cassisi \altaffilmark{1}}
\affil{Osservatorio Astronomico di Teramo, Via M. Maggini,
64100 Teramo, Italy; cassisi@astrte.te.astro.it}

\author{Vittorio Castellani \altaffilmark{2}}
\affil{Dipartimento di Fisica, Universit\`a di Pisa, Piazza Torricelli 2,
56100 Pisa, Italy; vittorio@astr1pi.difi.unipi.it}
\and
\author{Marcella Marconi}
\affil{Dipartimento di Fisica, Universit\`a di Pisa, Piazza Torricelli 2,
56100 Pisa, Italy; marcella@astr1pi.difi.unipi.it}

\altaffiltext{1}{Dipartimento di Fisica, Universit\`a de L'Aquila,
Via Vetoio, 67100 L'Aquila, Italy}

\altaffiltext{2}{Osservatorio Astronomico di Teramo, Via M. Maggini,
64100 Teramo, Italy}

\date{ }

\begin{abstract} 

\noindent
This paper presents evolutionary computations which investigate 
the theoretical predictions concerning metal rich RR Lyrae pulsators found 
both in the Galactic field and in the Galactic bulge. The main aim of this
investigation is to provide a homogeneous evolutionary context for further 
analyses concerning the pulsational properties of these evolving structures. 
In this connection a suitable set of stellar models characterized by two 
different metal contents, namely Z=0.01 and Z=0.02 were followed through 
both H and He burning phases. This evolutionary scenario covers the 
theoretical expectations for pulsating He burning structures with ages 
ranging from 20 to less than 1 Gyr. 

\noindent
For each given assumption about the 
star metallicity we find that "old" He burning pulsators with ages larger
than 2 Gyr have a common behavior, with Z=0.01 pulsators slightly 
more luminous and more massive than Z=0.02 pulsators. 
However, as soon as the metallicity increases above the solar value, 
the luminosity of Horizontal Branch (HB) stars increases again.
This effect is a direct consequence of the expected simultaneous increase 
of both original He and metals. 

\noindent
The occurrence of "young" RR Lyrae pulsators 
is discussed, and we show that this peculiar group of variable stars can  
be produced if the mass loss pushes He burning stars originated from more 
massive progenitors into the instability region where electron degeneracy 
plays a minor role. 
In this case the luminosity of the pulsators could be substantially
reduced with respect to the case of "old" variables, and therefore 
the theoretical expectations should also provide for this group of 
variable stars shorter pulsation periods. 

\noindent
Hydrogen burning isochrones for the quoted metallicity values are 
also presented and discussed. 

\noindent
Finally, the occurrence of a new "gravonuclear instability" in some HB 
models during the ignition of He shell burning is discussed.
We have found that the appearance of this phenomenon is due to the 
opacity "bump" connected with Iron. At the same time we have also 
outlined the role that this feature could play on observables. 
\end{abstract} 

\noindent
{\em Subject headings:} galaxies: stellar content -- globular clusters: 
general -- stars: evolution -- stars: horizontal branch -- stars: variables: 
other

\pagebreak
\section{INTRODUCTION}

\noindent
During the last decades  RR Lyrae pulsators in galactic globular clusters 
(GGC) have been the object
of lively debates and remarkable theoretical developments.
However, it has long been known that metal poor globular cluster
RR Lyrae are only a particular sample selected from a much
larger population for RR Lyrae pulsators in the Galaxy. Preston (1959)
was the first to draw attention to the group of metal rich RR Lyrae which
constitute a significant component ($\sim{25\%}$) of the RR Lyrae
population in the solar neighborhood. In order to estimate the metal
abundance of these objects he introduced the
$\Delta{S}$ parameter, i.e. the ratio between the spectral type based on 
Hydrogen Balmer lines and the spectral type based on the Calcium 
K lines. Due to the dependence of this spectral index on the pulsation 
phase, Preston suggested evaluating the $\Delta{S}$ 
parameter at minimum light and pointed out, for the first time, that RR Lyrae 
stars located in the solar neighborhood present a metal rich 
($0<\Delta{S}<2$) component, unknown in GGCs. At the same time this group 
of variable stars, in contrast with cluster variables, are also characterized 
by peculiar shorter periods.

\noindent
Calibration of $\Delta{S}$ parameter in terms of [Fe/H] (Preston 1961;
Butler 1975) has shown that this metal rich component reaches at least 
the solar metallicity, giving evidence for the occurrence
of RR Lyrae variables well beyond the canonical range for halo population
II stars. This observational evidence was further  strengthened
by Lub (1977), who presented an atlas of light and color curves
in the Walraven VBLUW photometric system for 90 RR Lyrae in the southern
field,  which confirms the occurrence of a metal rich component reaching a
solar-like composition. More recently, the investigation
of a sample of 302 nearby RR Lyrae stars  (Layden 1994,1995) has
disclosed a  metallicity distribution which
peaks at $[Fe/H]\approx{-1.5} $,  with a metal rich tail which extends to
$[Fe/H]\approx{0}$.

\noindent
The peculiar periods of  metal rich RR Lyrae represent a stimulating 
observational evidence to be connected with the evolutionary history of 
the pulsating structures. In a recent paper (Bono et al. 1996a and 
references therein) we have already shown that
a nonlinear hydrodynamical approach to stellar pulsations can be
connected with the prescriptions of stellar evolution for producing a
theoretical scenario throwing light on several observed features of
metal poor globular cluster  variables. The investigation will now be 
extended to metal rich pulsators, testing the capability of the
current theories to  account for the observed peculiar
pulsational behavior.

\noindent
This  paper deals with the results of an
evolutionary investigation devoted to deriving a theoretical scenario
for metal rich Galactic stellar populations. In a forthcoming paper
(Bono et al. 1996b) we will finally approach theoretical expectations 
concerning metal rich pulsators, investigating  the HR diagram location 
of the instability strip for both fundamental and first overtone modes
and discussing the pulsational properties of metal rich models
at selected luminosity levels. On this basis theoretical expectations
concerning the pulsational behavior of metal rich RR Lyrae will be
eventually compared with available data.

\section{LOW MASS STARS WITH SOLAR METALLICITY}

\noindent
According to a well-established scenario, RR Lyrae variables are 
unanimously recognized as HB stars which cross the instability strip during 
their central He burning evolutionary phase. In the case of globular 
clusters there is little doubt that RR Lyrae correspond to old He burning 
stars originated from Red Giant (RG) progenitors where electron degeneracy 
and neutrino cooling was efficient in the inner stellar structure
during giant branch evolution. 
As a consequence, we expect that the mass of the He core at the He ignition 
and the luminosity of the Zero Age Horizontal Branch (ZAHB) are only 
marginally dependent on the assumed cluster age. 

\noindent
However, unlike globular cluster stars, we have no clear 
constraints about the age of field variable stars and, in turn, we cannot 
exclude the occurrence of younger and therefore originally more massive 
structures, with rather different He cores and He burning luminosities. 
As a consequence, in order to provide a sound approach to the 
theoretical predictions of metal rich He burning models, a preliminary 
evaluation of both the He core masses at the He ignition ($M_{cHe}$) 
and of the amount of extra helium ($\Delta{Y}_{du}$) dredged up during 
the RG (H shell burning) evolutionary phase as a function of the 
stellar age over a suitable range of ages is necessary.

\noindent
For this purpose a detailed set of H burning models have been computed
following their evolution from the Zero Age Main Sequence (ZAMS) until 
the Helium ignition for two different values of the stellar metallicity, 
namely Z=0.01 and Z=0.02. In this section we present the results 
concerning both H and He burning phases for stars with solar metallicity,  
i.e. Z=0.02, extending previous computations of He burning models already 
in the literature references for this metal abundance down to ages lower 
than 1 Gyr. 
The next section presents similar computations for Z=0.01. These new 
sequences of models allow for a straight-forward analysis of the 
evolutionary behavior of moderately metal rich stellar populations 
which represent a link with more metal poor stars, rarely debated in
the literature.

\noindent
Stellar models have been computed adopting the latest version of the FRANEC 
evolutionary code (see Chieffi \& Straniero 1989), which includes several 
upgrades of the input physics. Major improvements are the opacity tables for 
stellar interiors provided by Roger \& Iglesias (1992) and low temperature 
molecular opacities for outer stellar layers by Alexander \& Ferguson (1994).
Both high and low temperature opacity tables assume Grevesse (1991)
solar chemical mixture. Solar metallicity models assume an initial 
amount of Helium as given by $Y=0.289$, whereas the mixing length parameter 
in the treatment of superadiabatic convective region  has been taken equal 
to 2.25 times the local pressure scale height ($H_p$).
According to the adopted physical inputs, these choices constrain the model 
with  mass, chemical composition and age of the Sun to fit solar luminosity 
and effective temperature, producing the so-called Standard Solar Model.
The adopted He abundance appears in good agreement with recent 
measurements of He abundance in the Galactic bulge which provide 
$Y=0.28\pm0.02$ (Minniti 1995).

\noindent
Table 1 gives selected evolutionary results for all the computed H burning 
models. The comparison of data in Table 1 with similar results already
in the literature (Castellani, Chieffi \& Straniero 1992; 
Horch, Demarque \& Pinsonneault 1992; Dorman, Rood \& O'Connell 1993;
Bressan et al. 1993;  Fagotto et al. 1994) shows reasonable agreement 
with marginal differences which
scarcely affect the evolutionary scenario. However, this is not 
the case for the more massive models in Bressan et al. (1993) and Fagotto 
et al. (1994) in which the assumption of efficient core overshooting 
sensibly affects the evolution of the structures.

\noindent
At the same time, the large number of evolutionary sequences computed until 
the Helium ignition allows for a detailed analysis of the "Red Giant 
Transition" (Sweigart, Greggio \& Renzini 1989) for a stellar population
with solar composition. Figure 1 shows the  mass  of the He core at the
He ignition, the luminosity at the Red Giant Branch (RGB) tip and  the
age of the models as function of the mass of the evolving star. We find 
that the transition occurs at the age 
$t\simeq{8.9}\ 10^8$ yrs, when at the tip of the RGB a star has  
$M\simeq2.15M_{\odot}$ and is characterized by a He core of 
$M_{cHe}\simeq0.41M_{\odot}$. For the sake of homogeneity
with previous works (Sweigart et al. 1989; Sweigart, Greggio \& Renzini 1990, 
hereinafter referred to as SGR; Cassisi \& Castellani 1993), we have 
defined the transition mass $M_{tr}$ as the stellar mass having a He core 
equal to the average value between the He core of degenerated structures 
and the absolute minimum in $M_{cHe}$ at the He ignition.
Comparison with previous results, as discussed in Cassisi \& Castellani 
(1996), discloses that present results predict rather smaller
transition masses (2.15 versus 2.6$M_{\odot}$) and therefore larger
transition ages (900 versus 500 Myr) than expected in the pioneering
paper by SGR 
but in agreement with previous computations by Castellani, Chieffi \& 
Straniero (1992). 

\noindent
As a whole, present
results confirm theoretical expectation that the age of low mass, 
metal rich RG stars, which experience strong electron degeneracy in the 
stellar core, can be much lower in comparison with metal poor stars. As a
consequence, at solar metallicity and for ages larger or of the order of
2.5 Gyr we expect a rather constant value for the mass of the He core
at the He flash and, in turn, a common luminosity level for the ZAHB 
models.

\noindent
These evolutionary results allow for the evaluation of updated 
H burning isochrones, which are presented in Figure 2 for selected 
assumptions about the stellar age. Table 2 summarizes the numerical 
data concerning the isochrones. This table
reports in column (1) the stellar age (Gyr), in column (2) the mass value
(solar units) of the model located at the Turn Off (TO) point, whereas 
in columns (3) and (4) are listed both the luminosity and the effective 
temperature of this point. Columns (5) and (6) present the V magnitude
and the B-V color of the TO point, respectively. Theoretical observables 
have been transformed into the observational plane according to the 
bolometric corrections and to the color-temperature relations provided by 
Kurucz (1992). We adopted a bolometric magnitude for the Sun of 
$M_{Bol,\odot}=4.75$ mag. The last column shows the V magnitude difference 
between the TO and the ZAHB models located at log $T_e=3.85$. 

\noindent
On the basis of these results the evolution of He burning models
has been investigated for selected assumptions about the cluster ages, 
i.e. for different masses of the RGB progenitors constructed by assuming 
different evolutionary values for both $M_{cHe}$ and $\Delta{Y}_{du}$. 
The stellar models selected as RGB
progenitors have masses $M/M_{\odot} = 0.8, 1.0, 1.5, 2.0, 2.1$. The age of
these models at the tip of the RGB ranges from  $28.7\times10^9$ yrs down to
 $9.34\times10^8$ yrs. For each given age various efficiencies of mass loss
during the RGB phase have been simulated by computing a set of He burning
models with the total mass decreasing from the original value (no mass loss)
down to, at least, 0.5 $M_{\odot}$. 
Figure 3 shows the evolutionary paths in the HR diagram for the "older" 
computed models, i.e. for models  with progenitor masses smaller than or 
equal to $1.5M_{\odot}$, whereas Figure 4 shows the evolutionary
path of HB stars with more massive progenitors. Table 3 reports the total 
stellar masses and the main evolutionary parameters of HB models 
for the two different assumptions on the chemical composition. 

\noindent
As expected, Figure 3 shows that for ages larger or of the order of 2.5 Gyr
the effect of ages on HB evolutionary structure can be largely neglected.
As discussed in the pioneering paper by Taam, Kraft \& Suntzeff (1976),
we find that when the metallicity is increased up to solar values the mass 
of ZAHB models at $\log{T_e}\simeq3.85$, i.e. in the expected region where 
HB stars experience radial pulsation instability, decreases to about 
$0.5M_{\odot}$. 
However, our models show a larger luminosity ($\log{L/L_{\odot}}\simeq1.52$ 
against 1.45) for the very simple reason that both our computations and the 
recent evaluations by Dorman et al. (1993) predict He core masses 
substantially larger than assumed in Taam et al. 
(1976, $M_{cHe}=0.45M_{\odot}$).
Figure 3 also shows that the adopted range of masses appears large enough
to cover the various different fates of low mass He burning stars, from the
AGB manqu\`e stars to structures that are massive enough to reach the 
thermal pulsating stage on the Asymptotic Giant Branch (AGB). 
The evolutionary properties of
old He burning stellar models with  solar metallicity  have been exhaustively
investigated  in recent times (Brocato et al. 1990; Castellani \& Tornamb\'e
1991; Horch et al. 1992; Dorman et al. 1993), therefore the global
characteristics are
well known and they will not be discussed here. However, let us underline
two remarkable features of the evolution which appear quite evident in
Figure 3:

\noindent
1) the location of the models along the ZAHB critically depends on the
envelope mass, so that a change in the envelope mass of only 
$0.03-0.04M_{\odot}$ shifts the ZAHB location from the low temperature 
region (${\log}T_e\simeq3.75$) to the high temperature region 
(${\log}T_e\simeq4.20$) which is located well beyond the blue side of the 
instability strip for RR Lyrae stars. As a consequence, HB stellar models 
with a solar metallicity show, depending on the envelope mass, a 
{\em flip - flop} behavior concerning the location on the ZAHB. 
As already discussed by Horch et al. (1992) and by Dorman et al. (1993), 
this peculiarity is a distinctive feature of old stellar populations with 
metallicity equal to or larger than the solar value. This of course leaves 
a small probability for the occurrence of HB radial pulsators;

\noindent
2) due to the enhanced efficiency of the H shell, during  central He  burning
the H-rich outer layers are efficiently burned into He. As a result, 
at the central He exhaustion metal rich HB stars present thinner H-rich 
envelopes in comparison with metal poor stars. Therefore the range of masses 
which show the typical AGB-manqu\'e morphology becomes larger 
(for example see Castellani et al. 1994).

\noindent
As a relevant point, Figure 4 shows that below 2.5 Gyr the location of HB
shows a non-negligible dependence on age assumptions. As matter of fact, 
if we take into account ages below 2.5 Gyr the mass of the He core becomes 
smaller. We consequently find that the luminosity level of the HB at 
$\log{T_e}= 3.85$ goes from
$\log L/L_{\odot}\simeq1.5$ for old (t $>$ 2.5 Gyr) stars down to
$\log L/L_{\odot}\simeq1.4$ for t=1.1 Gyr (2.0$M_{\odot}$ progenitor) and to 
$\log L/L_{\odot}\simeq1.3$ for t= 0.93 Gyr ($2.1 M_{\odot}$ progenitor).
According to data listed in Figure 1, the minimum luminosity of He burning 
stars should be reached for a progenitor as massive as M$\simeq2.3M_{\odot}$,
that is for He burning stars with ages of the order of 0.69 Gyr.
Numerical computations show that in such a case we expect stellar mass 
values around $0.36M_{\odot}$ and ZAHB luminosities of the order of 
$\log{L/L_{\odot}}\simeq 1.15$ for He burning stars located inside the 
instability strip ($\log{T_e}\simeq3.85$). 
Following such a theoretical evidence, present models foresee that 
possible "young" RR Lyrae should be characterized by both significantly 
lower luminosities and hence also by corresponding shorter periods.

\subsection{OCCURRENCE OF GRAVONUCLEAR INSTABILITIES}

\noindent
Figure 4  shows that at the onset of He shell burning some "young" models 
present the unexpected evidence of iterative runaways which cause 
a sudden change of the effective temperature and in turn of the AGB 
location of these models. Figure 5 shows that such an occurrence takes 
place only for less massive models which reach the AGB from hotter 
effective temperatures, whereas Figure 6 gives details
of the time behavior of both the luminosity and effective temperature of
the models during the runaway phase. The dynamics of these
"gravonuclear instabilities" is shown in Figure 7, in which we report the
time behavior of the various energy sources which rule the physical structure
of these models. Exhaustion of central He is followed, as expected, by the 
temporary reignition of the H burning shell. The classical evolutionary 
scenario predicts at this time that the reignition of the H burning shell
should be subsequently followed by the quiescent ignition of the He shell. 

\noindent
However, in the quoted models for the first time we find that the He shell 
ignition  causes a sudden and quick expansion of the structure which switches 
off the nuclear energy sources located both in the H and He shells. 
As soon as the external layers of the stellar envelope reach their maximum 
outward excursion, the structure experiences a contraction phase which implies 
both an increase of the effective temperature and the reignition 
of the H burning shell at first and later of the He burning shell. 
At this time the models have already performed a closed path in the HR 
diagram, and as soon as the reignition of He burning shell takes place they 
repeat the previous process with remarkable regularity.

\noindent
The physical mechanism which rules the loops performed by these models can 
be easily found in the large opacity of stellar envelopes characterized by 
solar metal content. In order to properly disentangle the effects introduced 
by metals on the envelope structure we have undertaken several numerical 
experiments by adopting an {\em ad hoc} metallicity distribution for the 
stellar layers located above the H burning shell.  
The models which present "gravonuclear loops" have been indeed stopped just 
before the ignition of the He shell, and the subsequent evolution has been 
computed by assuming, in the outermost regions located at temperatures lower
than $10^6$ K, the radiative opacity for a more metal poor mixture, 
namely  Y=0.28 and Z=0.0001. The evolutionary tracks plotted in Fig. 8 
undoubtedly show that "fictitious" track (dashed line), in contrast with 
standard track, does not experience the "gravonuclear loops" at the ignition 
of the He shell. Moreover, even though the two evolutionary sequences 
originate from a common initial condition the "fictitious" models evolve 
on a completely different path. 

\noindent
On the basis of this simple test we may conclude that the appearance of this
phenomenon in young metal rich stars is mainly due to the large opacity  
of the metal rich mixture and in particular to the "opacity bump"
produced by Iron in the temperature region located close to $2\times 10^5$ K
and for densities lower than $10^{-7}$ grcm$^{-3}$. Moreover, it is worth 
noting that in this temperature-density range the main opacity source of the
stellar envelope is given by the Iron "bump" i.e. the opacity peak 
due to this element is not only larger than the Helium peak but also
than the Hydrogen peak. The reader interested in a detailed discussion 
concerning this opacity feature is referred to the thorough papers  
recently provided by Iglesias et al. (1995); Iglesias \& Rogers (1996) and by 
Seaton et al. (1994).

\noindent
Due to the large opacity, the energy produced by the ignition of 
He shell burning cannot be easily dissipated through the large opacity
layers, producing an increase of the local temperature which further
enhances the energy production. In this first stage the mechanism is
quite similar to the He shell flash occurring in the normal thermal
pulses of AGB structures during the phase of He shell reignition. 
However, in the present case the expansion of the structures eventually 
quenches the process since the decrease of both temperature and density 
rapidly switches off the burning shells. Thus the process starts again and we 
find that the stars experience a few tens of similar "gravonuclear loops" 
before approaching a quiet He shell burning phase. As shown in Figure 6,
the stars  spend about 5 million years repeatedly crossing the instability
strip at two different luminosity levels, $\log{L/L_{\odot}}\simeq1.7$
and  1.9 respectively. Such an occurrence produces a small but not
negligible probability of observing similar "high luminosity" metal
rich type II Cepheid variables in the field (for en extensive 
discussion concerning this group of variable stars see Diethelm 1985).
In fact, the "young" HB model (\msun =0.50) which experiences the 
gravonuclear instability spends a time ranging from $\approx$ 13,000 yrs 
during the first loops to $\approx$ 17,000 yrs during the last ones inside 
the instability 
strip. These evolutionary times, when compared with the lifetime spent by 
this model in the phase of central He burning (t$\approx 1.9\times 10^8$ yrs)
provide a probability to detect a type II Cepheid which is roughly of the
order of $10^{-4}$. Therefore even though these radial pulsators are rare 
they could be properly identified in the huge photometric databases of 
variable stars recently collected by the microlensing experiments. 

\noindent
According to the above leading-term physical considerations, the more 
massive models which are located at lower effective temperatures do not 
experience a "gravonuclear instability" because the temperature gradient, 
due to the 
efficiency of convective transport, is smoothed out over the whole envelope. 
Therefore for these models the ignition of He shell burning takes place 
without the expansion of the external layers since the energy leakage 
through the envelope is governed by the convective motions. 

\noindent
As far as older models with larger He cores are concerned, the "gravonuclear 
loops" do not appear for a completely different reason. In fact, for these 
models the energy provided by the ignition of He shell burning cannot supply
the amount of energy necessary for the expansion of the envelope and 
hence for pushing an extended region of the stellar envelope in the 
temperature-density region where the opacities present the quoted "bump".  
Therefore for these models the key parameter which inhibits the appearance
of the "gravonuclear instability" is the ratio between the total stellar
mass and the envelope mass i.e. $q=M_{env}/M_{tot}$, where $M_{env}$ is 
the amount of mass located above the H shell. In fact this parameter  
attains too large values for stellar models located in this region of 
the HR diagram. 
A more quantitative analysis concerning the appearance of this phenomenon
and its dependence on astrophysical parameters such as metal content and the 
$q$ values will be discussed in a forthcoming paper (Bono et al. 1996c).  

\noindent 
Detailed information on the time evolution of He burning models is shown
in Figures 9 and 10, in which the time behavior of stellar 
luminosities and effective temperatures for all the computed models are
shown. As expected on very general grounds, "old" models with progenitors 
masses smaller than the RG transition mass (i.e. M$\leq$1.5 M$_{\odot}$) 
show quite similar central He burning lifetimes. The He burning lifetimes 
for younger models, characterized by larger values of the progenitor masses, 
increase following the decrease of the He core at the ignition of central 
He burning. The maximum He burning lifetime will be reached in correspondence 
of the minimum value of $M_{cHe}$, where models are expected to have 
comparable lifetimes in H and He burning phases.
In Figure 9 we can further appreciate the unusual large lifetimes of the
hotter, less massive models, which make these models good
candidates for UV emission in metal rich stellar clusters (see for example 
Greggio \& Renzini 1990; Dorman et al. 1993; Dorman, O'Connell \& Rood 1995).

\noindent
The results concerning He burning evolution can be finally combined with 
previous results for H burning stars to give the difference in luminosity 
between the HB and the  main sequence TO as a function of the cluster age, 
as reported in Table 2.

\section{LOW MASS STARS WITH A METAL CONTENT Z=0.01}

\noindent
As already discussed, in order to cover the metallicity which separates 
solar metallicity models from metal poor stellar models, the investigation
has been extended to a low metal abundance, namely Z=0.01. Assuming that the 
amount of original He is proportional to the heavy element content, 
according to a Helium enrichment ratio $\Delta Y/ \Delta Z \approx 3.0$, 
for these models we adopted Y=0.255. On the basis of the discussion
given in the previous sections, H-burning models for selected values
of the stellar mass have been followed up to the ignition of central
He burning. The masses of the evolved models have been again 
chosen such as to cover the RG transition phase. The dependence
of the He core mass on the stellar mass at the ignition of central He  
is reported in Figure 1 so that a proper comparison with similar 
data but for solar metallicity can be easily made. As already known (SGR),
we expect that the transition mass decreases as soon as the metallicity 
is decreased and/or the amount of original He is increased.   
The evidence that our Z=0.01 models undergo the transition at masses 
similar to  the solar case suggests that the difference in the initial 
Helium abundance is counteracted by the variation in metallicity. 

\noindent
Selected data for H-burning models are reported in Table 4.
Comparison between data in Table 4 and similar data given in Table
1 for the solar metallicity case shows that when the metallicity 
decreases from the solar case down to Z=0.01 the transition mass 
increases from $M_{tr}=2.15M_{\odot}$ to about
$M_{tr}=2.21M_{\odot}$, whereas the transition age slightly decreases.
Figure 11 shows a selected sample of H-burning isochrones covering the
age range from 900 Myr to 20 Gyr. Numerical data concerning these isochrones
are given in Table 5, in the same order and with the
same content as already given in Table 2 for solar metallicity
stars. For the given original chemical composition we find that "old"
RG stars with electronically degenerate He cores ignite 3$\alpha$
reactions when the central He core has reached the mass 
$M_{cHe} \approx0.485M_{\odot}$ and the H-rich envelope has been enriched 
in Helium by ${\Delta}Y_{du}\simeq0.02$. The amount of Helium dredged up 
slightly depends on the age of the giant at the He flash. It is noteworthy 
that the decrease in both Z and Y acts in the direction of increasing 
the core mass at the He flash, as suggested by the computations.

\noindent
Structural parameters given by H burning evolution have been adopted 
to produce ZAHB models which have been further followed all along their 
He burning evolutionary phase. The evolutionary parameters adopted for 
these He burning models are listed in Table 3. 
Figure 12 shows the path in the HR diagram of selected
models  as originated from a 1.0M$_{\odot}$ progenitor,
which can be regarded as representative of the common behavior
of "old" He burning objects originated from RG stars with electronically
degenerate cores.  
A glance at the evolutionary tracks plotted in this figure shows that 
the "gravonuclear loops" now appear at the end of the He shell burning 
phase. These models do not play any role for properly defining 
the physical structure of RR Lyrae variables since they are located well
behind  the first overtone blue edge and therefore they will not be 
discussed here further. 

\noindent
However, the location in the HR diagram of the 
models which experience the "gravonuclear instabilities" before approaching 
the White Dwarf cooling sequence presents an intriguing feature worth 
being discussed. In fact, Figure 12 shows that the occurrence of the 
"gravonuclear loops" takes place at an effective temperature higher than  
$\approx 25,000$ K. As a consequence this mechanism could provide  
useful constraints on the appearance of the UV upturn observed in elliptical 
galaxies (Ferguson et al. 1991; Ferguson 1995 and references therein) 
which is approximately located around the quoted value of effective 
temperature. At the same time the occurrence of "gravonuclear instabilities"
provides a plain physical justification for the plausible existence of a
"critical metallicity" value, i.e. the metal abundance above which the HB 
morphology and in particular the evolution of hot HB and AGB manqu\`e 
stars is influenced by this parameter (for a critical theoretical review  
see Greggio \& Renzini 1990; Brocato et al. 1990; Dorman et al. 1995
and references therein). 
This region of the HR diagram should be investigated in more detail since 
it turns out that these evolutionary phases can provide important
clues concerning the final fate of low mass stars. 

\noindent
Comparison with results given in the previous section 
discloses than in the case Z=0.01 the ZAHB location appears moderately more
luminous than in the solar case. At $\log{T_e}=3.85$ we find
now $\log{L/L_{\odot}}=1.55$ against $\log{L/L_{\odot}}=1.52$ for Z=0.02.
It turns out that such a difference is governed by the variation of
the metal content, since the decrease in Y alone would produce fainter HB. 
These two evaluations of the ZAHB luminosity can be integrated
with similar data computed with exactly the same evolutionary 
code and physical inputs but for lower metallicity by Cassisi \& Salaris 
(1996 and references therein) to produce an updated evaluation of the ZAHB 
luminosity as a function of the metallicity covering the range 
from Z=0.0001 to Z=0.02.
Numerical values concerning such a parameter are reported in
Table 6 and are displayed in Figure 13, together with other
evolutionary parameters  relevant for He burning models.

\noindent
As already discussed in Castellani, Chieffi \& Pulone (1991 hereinafter 
referred to as CCP), we find that the relation $\log{L(3.85)}$ 
(i.e. the ZAHB luminosity at the effective temperature $\log{T_e}=3.85$)
versus $\log{Z}$ presents a rather constant slope in the
interval Z=0.0001-0.001, whereas for metallicity larger than
Z=0.001 the dependence of $\log{L(3.85)}$ on the metallicity progressively
increases. Nevertheless, Figure 13 now shows that above Z=0.006 
the quoted dependence decreases again. Comparison with similar data given
in CCP but for a constant value of the original He (Y=0.23) 
convincingly demonstrates that such a feature must be regarded
as an evidence of the fact that the increase in He connected with the 
increase in metallicity starts to play a relevant role in determining the
ZAHB luminosity. On the basis of the assumed Helium enrichment ratio, 
we can therefore suspect that for even larger metallicities
the increase in He will eventually dominate, reversing the dependence 
of the HB on the content of heavy elements. This hypothesis has been 
tested by computing selected stellar models with a metal abundance a 
factor of two larger than the solar one. The result, as reported in both
Table 6 and figure 13, supports such an occurrence and shows that the 
ZAHB luminosity reaches a minimum for solar composition,
whereas for metallicity larger than this value it presents an upturn.  

\noindent  
Data concerning the ZAHB luminosity can be translated in absolute
V magnitude according to Kurucz's (1992) atmosphere models. On the 
basis of the data displayed in Figure 14, we find that the 
$\log{L(3.85)}$, in the lower metallicity range (Z= 0.0001 - 0.001), 
scales with metallicity according to the following relation:
 
\begin{center}
$M_{V}(3.85)= 0.16\cdot{\log}Z + 1.19 \approx 0.16\cdot{[Fe/H]} + 0.91$
\end{center}

\noindent 
which appears in excellent agreement with observational estimate of the 
slope of this dependence as given by Walker (1992) and Carney, Storm 
\& Jones (1992), i.e.  $\Delta{M_V}=0.15\cdot{\Delta}[Fe/H]$ 
and with a zero point which appears intermediate between the values given 
in the two above quoted investigations.
The same Figure shows that when the metallicity is increased up to values  
as large as 0.006, the linear relation connecting the extreme lower limit 
takes the form:

\begin{center}
$M_V(3.85)= 0.21\cdot{\log}Z + 1.36 \approx 0.21\cdot{[Fe/H]} + 1.00$
\end{center}

\noindent
with a much larger "mean" dependence of HB magnitudes on the
metallicity, i.e. as large as $\Delta{M_V}=0.21\cdot{\Delta}[Fe/H]$.
However, it is worth noting that the zero point of such relations
depends on both the adopted bolometric corrections and the bolometric
magnitude assumed for the Sun.
 
\noindent
The effect of lowering the age below the time of the RG
transition is shown in Figure 15, which presents the evolutionary
paths of a He burning model originated from a progenitor of
2.0$M_{\odot}$ and hence for an age of about 0.98 Gyr. 
Inspection of this figure shows that
the gravonuclear instabilities have now disappeared,
following the decreased opacity of the stellar envelopes.
Comparison with the HR diagram location of "old" He burning models
for the same value of Z shows that again when the age becomes 
lower the HB stars can sensibly attain lower luminosities
($\log{L/L_{\odot}}\simeq1.46$).
Again the minimum luminosity will be reached in correspondence of
the minimum value of $M_{cHe}$, namely for a mass $M\approx2.3M_{\odot}$
at the age $\approx0.63$ Gyr. Calculations show that in such a case
at $\log{T_e}=3.85$ we should find He burning models with mass value 
$M\approx0.36M_{\odot}$ and with ZAHB luminosity as low as 
$\log{L(3.85)}\simeq 1.16$.

\section{DISCUSSION AND CONCLUSIONS}

\noindent
In this paper we have explored the theoretical scenario 
concerning the evolutionary status of metal rich RR Lyrae pulsators
in order to make theoretical predictions about the astrophysical  
parameters governing the pulsation phenomenon, namely stellar mass and 
luminosity. In order to provide a reliable and homogeneous scenario 
for these predictions, evolutionary
computations covering both H and He burning phases have been presented
for selected choices about the original mass of the evolving 
stars and for two different assumptions about the star metallicity, namely
Z=0.01 and 0.02.

\noindent
When dealing with "old" HB stars originated from
RG progenitors with electronically degenerate stellar cores
we find that above Z=0.006 the HB luminosity keeps decreasing when 
the metallicity increases, until it reaches a luminosity minimum
at solar metallicity values. In more metal rich stars the increase 
in the original He abundance should dominate the evolution, 
increasing again the luminosity of HB structures. Due to the already 
known (CCP) nonlinear behavior of the ZAHB luminosity with the metallicity 
it turns out that the $M_V(HB) - [Fe/H]$ relations based on metal poor 
globular cluster stars (Carney, Storm \& Jones 1992; Walker 1992; 
Sandage 1993) cannot be safely extrapolated at higher metallicities. 
As a consequence even though the absolute magnitude of RR Lyrae itself, 
which is characterized by a metal abundance close to the solar value, 
has been recently provided by the Hipparcos mission (Perryman et al. 1995) 
this firm observational constraint cannot be {\em tout court} adopted to 
infer the reliability of the quoted relations in the metal poor range. 

\noindent
The occurrence of "young" RR Lyrae pulsators can be produced if mass 
loss pushes into the instability region He burning 
stars originated from more massive progenitors in which electron 
degeneracy plays a reduced or a negligible role. 
In this case the luminosity of the pulsators could be substantially
reduced with respect to the case of "old" variables, thus decreasing
theoretical expectations on pulsational periods. 

\noindent
We have found an interesting new "gravonuclear instability" connected with 
the high opacity of metal rich stellar envelopes of HB stars. The appearance 
of this feature can be understood in terms of leading-term physical 
arguments and the role played by the Iron opacity bump. 
On the basis of both luminosity and temperature excursions experienced
by the solar metallicity models during the "gravonuclear loops" we 
predict that pulsation properties of metal rich Type II Cepheids could 
provide useful clues concerning the appearance of this phenomenon. 
Moreover, we have found that the "gravonuclear loops" in HB models 
characterized by a metal content of Z=0.01 take place in a different 
region of the HR diagram in comparison with the solar metallicity models. 
In fact, for the former models the "gravonuclear instabilities" are located 
at both higher luminosities and higher effective temperatures. 
This new finding adds valuable pieces of information for understanding 
the UV upturn observed in elliptical galaxies and offers a coherent picture
concerning the role played by a "critical metallicity" value on the 
evolution of hot HB stars.  

\noindent
This general evolutionary scenario is presented to allow for  
an extensive approach to the pulsational properties of metal rich
RR Lyrae stars, but also for disclosing the astrophysical parameters
which govern the evolutionary properties of field stars belonging 
to both the Galactic bulge and the solar neighborhood. 

\noindent
It is a pleasure to thank B. Dorman for many stimulating and valuable 
discussions on the advanced evolutionary phases of low mass stars. We are 
also grateful to an anonymous referee for the pertinence of his/her
comments regarding the content and the style of an early draft of this 
paper which improved its readability. This research has made use of 
NASA's Astrophysics Data System Abstract Service. This work was partially 
supported by MURST, CNR-GNA and ASI.


\pagebreak

\section{Figure Captions}

\vspace*{3mm} \noindent {\bf Fig. 1.} Evolutionary parameter for structure
at the He ignition as a function of the stellar mass. From the top to the 
bottom are plotted the stellar age, the luminosity and the mass of the He 
core.

\vspace*{3mm} \noindent {\bf Fig. 2.} Theoretical isochrones for the H
burning phases of solar metallicity structures, for the labeled assumptions
about the cluster age. The time interval between consecutive isochrones is 
1 Gyr, with the exception of the two isochrones corresponding to 0.9 Gyr 
and 1 Gyr respectively.

\vspace*{3mm} \noindent {\bf Fig. 3.} Comparison of the evolutionary paths 
in the HR diagram of "old" He burning structures. The squares mark the ZAHB 
location of two models computed by adopting total stellar masses of 
\msun =0.52 and 0.54 respectively, and a RG progenitor mass of \msun =0.8. 
The stellar mass values of the progenitors and the He core masses are 
labeled. The total stellar masses adopted in each set of HB models are 
reported in Table 3. 

\vspace*{3mm} \noindent {\bf Fig. 4.} Evolutionary paths in the HR diagram 
of "young" He burning structures compared with similar paths but for
"old" models. Symbols are the same as in Fig. 3. The stellar mass values of 
the progenitors and the He core masses are labeled. The total stellar masses 
adopted in each set of HB models are reported in Table 3. 

\vspace*{3mm} \noindent {\bf Fig. 5.} The evolutionary tracks for the two 
models which experience "gravonuclear" instabilities. For these models the 
mass of the progenitor is equal to  $2.0M_{\odot}$.

\vspace*{3mm} \noindent {\bf Fig. 6.} {\em Top panel}: The time behavior 
of the surface luminosity and effective temperature during the 
"gravonuclear loops" of the $0.50M_{\odot}$ model. {\em Bottom panel}: 
same as top panel but referred to the $0.49M_{\odot}$ model.

\vspace*{3mm} \noindent {\bf Fig. 7.} Time behavior of the gravitational
luminosity ({em panel a}), of the hydrogen ({\em panel b}) and helium 
({\em panel c}) luminosities during the "gravonuclear loops" for the 
$0.50M_{\odot}$ model. 

\vspace*{3mm} \noindent {\bf Fig. 8.} Comparison between the standard
evolutionary track (solid line) and the "fictitious" sequence of models 
(dashed line) constructed by adopting for the stellar layers located above 
the H burning shell ($T \le 10^6$ K) the radiative opacities of 
a "fictitious" metal poor (Y=0.28, Z=0.0001) mixture. The circle marks the 
evolutionary phase which precedes the ignition of the He shell burning. 
See text for further explanation.

\vspace*{3mm} \noindent {\bf Fig. 9.} Time behavior of the luminosity for 
all He burning models with solar metallicity. Physical parameters are 
labeled.

\vspace*{3mm} \noindent {\bf Fig. 10.} Time behavior of the effective 
temperature for all He burning models with solar metallicity. Physical 
parameters are labeled. 

\vspace*{3mm} \noindent {\bf Fig. 11.} Same as Figure 2, but referred to 
a different chemical composition, namely  Z=0.01 and Y=0.255.

\vspace*{3mm} \noindent {\bf Fig. 12.} Evolutionary path in the HR diagram 
of selected He burning models computed by assuming a progenitor mass value
of $M = 1.0M_{\odot}$ and a metallicity Z=0.01. 
The He core mass is labeled, whereas the total stellar masses adopted are
reported in Table 3.

\vspace*{3mm} \noindent {\bf Fig. 13.} {\em Panel a}: Helium abundance in 
the envelope as a function of metallicity for the $M=0.8M_{\odot}$ model 
at the RGB tip. The adopted initial Helium abundance -dashed line- is also 
shown. {\em Panel b}: the mass of the He core at the He burning ignition 
for the same model, whereas {\em panel c} shows the luminosity of the ZAHB 
at the effective temperature $\log{T_e}=3.85$ for an "old" Red Giant 
progenitor. See text for further explanation.

\vspace*{3mm} \noindent {\bf Fig. 14.} The V magnitude of ZAHB at 
$\log{T_e}=3.85$ as a function of metallicity for an "old" Red Giant 
progenitor.  The bolometric corrections given by Kurucz (1992) have been 
used. We also adopted a bolometric magnitude for the Sun equal to 4.75 mag.

\vspace*{3mm} \noindent {\bf Fig. 15.} Same as as Figure 12, but the 
evolutionary tracks have been constructed by assuming an initial 
mass value for the progenitor equal to  $2.0M_{\odot}$.
In order to avoid crowding problems the evolutionary paths located 
at the higher temperatures have been plotted by adopting dotted and 
dashed lines. The He core mass is labeled, whereas the total stellar 
masses adopted are reported in Table 3.

\clearpage
\begin{deluxetable}{ccccc}
\footnotesize  
\tablecaption{Selected theoretical evolutionary quantities for the models 
with solar metallicity.}\label{tbl-1}
\tablehead{
\colhead{$M/M_\odot$\tablenotemark{a}}&
\colhead{$\log(L/L_\odot)_{tip}$\tablenotemark{b}}& 
\colhead{$\log{t_{He}}$\tablenotemark{c}}&
\colhead{$M_{cHe}$\tablenotemark{d}}&
\colhead{${\Delta}Y_{du}$\tablenotemark{e}} }
\startdata
  0.8	&   3.442 & 	10.458 &   0.479 & 0.017 \nl
  1.0   &   3.439 &	10.095 &   0.478 & 0.021 \nl	
  1.2	&   3.434 &	 9.804 &   0.477 & 0.019 \nl
  1.4   &   3.433 &      9.565 &   0.477 & 0.016 \nl
  1.5	&   3.433 &	 9.462 &   0.476 & 0.014 \nl
  1.6   &   3.430 &      9.363 &   0.476 & 0.013 \nl
  1.8   &   3.392 &      9.185 &   0.469 & 0.010 \nl
  2.0	&   3.255 &	 9.036 &   0.445 & 0.009 \nl
  2.1   &   3.128 &      8.970 &   0.425 & 0.007 \nl
  2.3   &   2.393 &      8.840 &   0.339 & 0.011 \nl
  2.4   &   2.473 &      8.785 &   0.342 & 0.011 \nl
  2.5   &   2.487 &      8.733 &   0.345 & 0.012 \nl
  3.0   &   2.587 &      8.509 &   0.391 & 0.015 \nl
\enddata
\tablenotetext{a}{Stellar mass value (solar units).}
\tablenotetext{b}{Logarithmic luminosity (solar units) at the RGB tip.}
\tablenotetext{c}{Logarithmic age (yrs) at the RGB tip.}
\tablenotetext{d}{Mass value of the He core (solar units) at the He ignition.}
\tablenotetext{e}{Amount of extra-helium dredged up during RGB evolution}
\end{deluxetable}

\clearpage
\begin{deluxetable}{ccccccc}
\footnotesize  
\tablecaption{Selected evolutionary data concerning the solar metallicity isochrones.}\label{tbl-2}
\tablehead{
\colhead{Age (Gyr)}&
\colhead{$M/M_{\odot}$\tablenotemark{a}}& 
\colhead{$\log(L/L_\odot)$\tablenotemark{b}}&
\colhead{$\log{T_e}$\tablenotemark{c}}& 
\colhead{$M_V$\tablenotemark{d}}& 
\colhead{$(B-V)$\tablenotemark{e}}& 
\colhead{${\Delta}V^{TO}_{HB}$\tablenotemark{f}} }
\startdata
  0.9 & 2.049 & 1.500 & 3.910 & 1.080 &  0.139 & -0.39\nl
  1.0 & 1.976 & 1.435 & 3.899 & 1.233 &  0.177 & -0.15\nl
  2.0 & 1.566 & 1.018 & 3.835 & 2.307 &  0.390 &  1.29\nl
  3.0 & 1.372 & 0.779 & 3.815 & 2.922 &  0.461 &  1.95\nl
  4.0 & 1.255 & 0.606 & 3.802 & 3.367 &  0.510 &  2.39\nl
  5.0 & 1.173 & 0.472 & 3.791 & 3.714 &  0.553 &  2.74\nl
  6.0 & 1.118 & 0.385 & 3.782 & 3.944 &  0.588 &  2.97\nl
  7.0 & 1.072 & 0.316 & 3.775 & 4.127 &  0.616 &  3.15\nl
  8.0 & 1.050 & 0.275 & 3.770 & 4.237 &  0.637 &  3.26\nl
  9.0 & 1.020 & 0.240 & 3.766 & 4.332 &  0.653 &  3.36\nl
 10.0 & 0.996 & 0.203 & 3.762 & 4.432 &  0.669 &  3.46\nl
 11.0 & 0.975 & 0.170 & 3.758 & 4.523 &  0.686 &  3.55\nl
 12.0 & 0.949 & 0.140 & 3.754 & 4.606 &  0.703 &  3.63\nl
 13.0 & 0.930 & 0.112 & 3.750 & 4.685 &  0.719 &  3.72\nl
 14.0 & 0.916 & 0.089 & 3.747 & 4.750 &  0.732 &  3.78\nl
 15.0 & 0.902 & 0.072 & 3.744 & 4.800 &  0.744 &  3.82\nl
 16.0 & 0.891 & 0.058 & 3.742 & 4.840 &  0.752 &  3.87\nl
 17.0 & 0.880 & 0.049 & 3.739 & 4.871 &  0.765 &  3.90\nl
 18.0 & 0.870 & 0.037 & 3.736 & 4.909 &  0.777 &  3.93\nl
 19.0 & 0.858 & 0.025 & 3.734 & 4.945 &  0.785 &  3.97\nl
 20.0 & 0.847 & 0.010 & 3.732 & 4.989 &  0.793 &  4.02\nl
\enddata
\tablenotetext{a}{Mass value at the TO point (solar units).}
\tablenotetext{b}{Logarithmic luminosity of the TO point (solar units).}
\tablenotetext{c}{Logarithmic effective temperature (K) of the TO point.}
\tablenotetext{d}{V Magnitude of the TO point (mag).}
\tablenotetext{e}{B-V color of the TO point (mag).}
\tablenotetext{f}{V Magnitude difference between the TO point and the ZAHB at $\log T_e=3.85$.}
\end{deluxetable}

\clearpage
\begin{deluxetable}{cccc}
\footnotesize  
\tablecaption{Selected evolutionary parameters for the He burning phase for 
different assumptions on chemical composition and RG progenitor masses.}\label{tbl-3}
\tablehead{
\colhead{$M_{pr}$\tablenotemark{a}}&
\colhead{$Y_{HB}$\tablenotemark{b}}&
\colhead{$M_{cHe}$\tablenotemark{c}}&
\colhead{$M_{tot}$\tablenotemark{d}} }
\startdata
     \multicolumn{4}{c}{Z=0.02 ~~~~~$Y_{ZAMS}=0.289$}\nl 
 0.8 & 0.306 & 0.479 & 0.490, 0.500, 0.510, 0.520, 0.530, 0.540\nl
     &       &       & 0.550, 0.600, 0.650, 0.700, 0.750, 0.800\nl
 1.0 & 0.310 & 0.478 & 0.490, 0.500, 0.550, 0.600, 0.650, 0.700\nl
     &       &       & 0.750, 0.800, 0.850, 0.900, 0.950, 1.000\nl
 1.5 & 0.303 & 0.477 & 0.500, 0.550, 0.600, 0.650, 0.700, 0.750\nl
     &       &       & 0.800, 0.900, 1.000, 1.100, 1.200, 1.300\nl
     &       &       & 1.400, 1.500\hspace*{40.0mm} \nl
 2.0 & 0.298 & 0.445 & 0.490, 0.500, 0.510, 0.520, 0.550, 0.600\nl
     &       &       & 0.900, 1.000, 1.200, 1.300, 1.400, 1.500\nl
     &       &       & 1.600, 1,800, 1.900, 2.000\hspace*{20.0mm}\nl
 2.1 & 0.296 & 0.425 & 0.500, 1.000, 1.200, 1.400, 1.600, 1.800\nl
     &       &       & 2.000, 2.100\hspace*{40.0mm} \nl
     &	     &       & \vspace*{-2.0mm}\nl
     \multicolumn{4}{c}{Z=0.01 ~~~~~$Y_{ZAMS}=0.255$} \vspace*{-2.0mm}\nl 
         &	      &       & \nl
 1.0 & 0.277 & 0.487 & 0.495, 0.500, 0.510, 0.520, 0.530, 0.540\nl
     &       &       & 0.550, 0.560, 0.570, 0.580, 0.590, 0.600\nl
     &       &       & 0.630, 0.650, 0.700, 0.750, 0.800, 0.900\nl
     &       &       & 1.000\hspace*{50.0mm}\nl
 2.0 & 0.266 & 0.464 & 0.465, 0.470, 0.475, 0.480, 0.490, 0.500\nl
     &       &       & 0.510, 0.520, 0.530, 0.540, 0.550, 0.580\nl
     &       &       & 0.600, 0.700, 0.800, 1.000, 1.200, 1.600\nl
     &       &       & 1.800, 2.000\hspace*{40.0mm}\nl
\enddata
\tablenotetext{a}{ Stellar mass value for the RG progenitor (solar units).  
\hspace*{0.5mm} $^b$ Helium abundance adopted for HB models.  
\hspace*{0.5mm} $^c$ Helium core mass value (solar units). 
\hspace*{0.5mm} $^d$ Total stellar mass value adopted for HB models (solar units).}  
\end{deluxetable}

\clearpage
\begin{deluxetable}{ccccc}
\footnotesize  
\tablecaption{Same as Table 1 but the data are referred to a different 
chemical composition: Z=0.01 and Y=0.255.}\label{tbl-4}
\tablehead{
\colhead{$M/M_\odot$} & \colhead{$\log(L/L_\odot)_{tip}$} & 
\colhead{$\log{t_{He}}$} & \colhead{$M_{cHe}$} & \colhead{${\Delta}Y_{du}$} }
\startdata
 0.8 &  3.447 &  10.402 & 0.489 & 0.017 \nl
 1.0 &  3.442 &  10.038 & 0.487 & 0.022 \nl
 1.2 &  3.438 &   9.745 & 0.485 & 0.023 \nl
 1.5 &  3.435 &   9.410 & 0.485 & 0.017 \nl
 1.6 &  3.434 &   9.315 & 0.485 & 0.015 \nl
 1.8 &  3.418 &   9.140 & 0.482 & 0.012 \nl
 2.0 &  3.323 &   8.991 & 0.464 & 0.011 \nl
 2.2 &  3.088 &   8.866 & 0.426 & 0.012 \nl
 2.3 &  2.434 &   8.798 & 0.331 & 0.013 \nl
 2.5 &  2.499 &   8.695 & 0.341 & 0.014 \nl
 3.0 &  2.562 &   8.481 & 0.381 & 0.017 \nl
\enddata
\end{deluxetable}

\clearpage
\begin{deluxetable}{ccccccc}
\footnotesize  
\tablecaption{Same as Table 2 but the data are referred to a different 
chemical composition: Z=0.01 and Y=0.255.}\label{tbl-5}
\tablehead{
\colhead{$Age$} & \colhead{$M/M_{\odot}$} & \colhead{$\log(L/L_\odot)$} &
\colhead{$\log{T_e}$} & \colhead{$M_V$} & \colhead{$(B-V)$} & \colhead{${\Delta}V^{TO}_{HB}$} }
\startdata
 0.9 & 1.748 & 1.144 & 3.903 & 1.993 &  0.175 & 0.82 \nl
 1.0 & 1.686 & 1.081 & 3.892 & 2.149 &  0.204 & 0.98 \nl
 2.0 & 1.490 & 1.012 & 3.863 & 2.333 &  0.293 & 1.39 \nl
 3.0 & 1.297 & 0.736 & 3.834 & 3.043 &  0.379 & 2.10 \nl
 4.0 & 1.184 & 0.561 & 3.814 & 3.496 &  0.441 & 2.55 \nl
 5.0 & 1.129 & 0.489 & 3.804 & 3.683 &  0.477 & 2.74 \nl
 6.0 & 1.085 & 0.432 & 3.796 & 3.833 &  0.504 & 2.89 \nl
 7.0 & 1.047 & 0.379 & 3.790 & 3.971 &  0.526 & 3.03 \nl
 8.0 & 1.016 & 0.338 & 3.785 & 4.079 &  0.545 & 3.14 \nl
 9.0 & 0.990 & 0.306 & 3.781 & 4.164 &  0.562 & 3.22 \nl
10.0 & 0.965 & 0.272 & 3.776 & 4.256 &  0.579 & 3.31 \nl
11.0 & 0.944 & 0.243 & 3.772 & 4.332 &  0.594 & 3.39 \nl
12.0 & 0.925 & 0.218 & 3.768 & 4.401 &  0.609 & 3.46 \nl
13.0 & 0.908 & 0.194 & 3.765 & 4.465 &  0.622 & 3.52 \nl
14.0 & 0.890 & 0.163 & 3.762 & 4.549 &  0.635 & 3.61 \nl
15.0 & 0.876 & 0.143 & 3.759 & 4.604 &  0.646 & 3.66 \nl
16.0 & 0.862 & 0.116 & 3.756 & 4.676 &  0.657 & 3.73 \nl
17.0 & 0.849 & 0.091 & 3.754 & 4.744 &  0.667 & 3.80 \nl
18.0 & 0.837 & 0.068 & 3.751 & 4.806 &  0.677 & 3.87 \nl
19.0 & 0.827 & 0.054 & 3.749 & 4.846 &  0.685 & 3.91 \nl
20.0 & 0.817 & 0.034 & 3.747 & 4.902 &  0.694 & 3.96 \nl
\enddata
\end{deluxetable}

\clearpage
\begin{deluxetable}{ccccccc}
\footnotesize  
\tablecaption{Selected evolutionary parameters at the RGB tip and at the 
ZAHB for different assumptions concerning the metallicity. All data are 
referred to an "old" cluster population.}\label{tbl-6} 
\tablehead{
\colhead{$Z$\tablenotemark{a}}& 
\colhead{$Y_{ZAMS}$\tablenotemark{b}}& 
\colhead{$Y_{ZAHB}$\tablenotemark{c}}& 
\colhead{$M_{cHe}$\tablenotemark{d}}& 
\colhead{$\log{L(3.85)}$\tablenotemark{e}}& 
\colhead{$M(3.85)/M_{\odot}$\tablenotemark{f}}& 
\colhead{$M^{ZAHB}_V$\tablenotemark{g}} }
\startdata
  0.0001 & 0.230 & 0.238 & 0.511 &  1.735 &  0.796 &  0.557 \nl 
  0.0003 & 0.230 & 0.240 & 0.507 &  1.699 &  0.711 &  0.636 \nl
  0.0006 & 0.230 & 0.242 & 0.504 &  1.678 &  0.671 &  0.679 \nl     
  0.0010 & 0.230 & 0.243 & 0.502 &  1.661 &  0.648 &  0.715 \nl  
  0.0030 & 0.230 & 0.245 & 0.498 &  1.610 &  0.608 &  0.823 \nl 
  0.0060 & 0.230 & 0.246 & 0.495 &  1.556 &  0.585 &  0.941 \nl
  0.0100 & 0.255 & 0.272 & 0.489 &  1.542 &  0.575 &  0.962 \nl
  0.0200 & 0.289 & 0.306 & 0.479 &  1.523 &  0.545 &  0.973 \nl
  0.0400 & 0.340 & 0.357 & 0.468 &  1.533 &  0.540 &  0.938 \nl
\enddata
\tablenotetext{a}{Metal content.} 
\tablenotetext{b}{Initial He abundance on the Zero Age Main Sequence.} 
\tablenotetext{c}{He abundance in the envelope of the ZAHB models.} 
\tablenotetext{d}{Mass value of He core at the RGB tip (solar units).} 
\tablenotetext{e}{Logarithmic ZAHB luminosity taken at $\log{T_e}=3.85$.} 
\tablenotetext{f}{Mass value of the ZAHB model located at $\log T_e=3.85$.} 
\tablenotetext{g}{V magnitude of the ZAHB.} 
\end{deluxetable}

\end{document}